\documentclass[11pt]{article}
\usepackage{moriond,epsfig}

\def\aap{A\&A}
\def\apjl{ApJL}
\def\apj{ApJ}
\def\mnras{MNRAS}

\def\be{\begin{equation}}
\def\ee{\end{equation}}
\def\bea{\begin{eqnarray}}
\def\eea{\end{eqnarray}}

\begin{document}
\title{STUDYING THE EVOLUTION OF MULTI-WAVELENGTH EMISSIVITIES WITH THE VIMOS VLT DEEP SURVEY}
 
\author{L. TRESSE$^1$, O. ILBERT$^{1,2}$, E. ZUCCA$^2$, G. ZAMORANI$^2$, S. ARNOUTS$^1$, S. BARDELLI$^2$ \\ AND THE VVDS TEAM$^\dag$\footnote[0]{$^\dag$D. Bottini (INAF-IASF Milano) , B. Garilli  (INAF-IASF Milano), V. Le Brun (LAM  Marseille), O. Le F\`evre (LAM Marseille), D. Maccagni  (INAF-IASF Milano), J.-P. Picat (LAOMP Toulouse), R. Scaramella, M. Scodeggio  (INAF-IASF, Milano), G. Vettolani (INAF-IRA Bologna), A. Zanichelli (INAF-IRA Bologna), C. Adami (LAM Marseille), M. Bolzonella (INAF-OA Bologna), A. Bongiorno (Univ. Bologna), A. Cappi (INAF-OA Bologna), S. Charlot (IAP Paris), P. Ciliegi  (INAF-OA Bologna), T. Contini (LAOMP Toulouse),  O. Cucciati (INAF-OA Brera), S. Foucaud (Univ. Nottingham), P. Franzetti  (INAF-IASF, Milano), I. Gavignaud (AIP Potsdam), L. Guzzo (INAF-OA Brera), A. Iovino (INAF-OA Brera), F. Lamareille  (LAOMP Toulouse), H.J. McCracken (IAP Paris), B. Marano (Univ. Bologna), C. Marinoni (CPT Marseille), A. Mazure (LAM Marseille), B. Meneux (INAF-OA Brera), R. Merighi (INAF-OA Bologna), S. Paltani (ISDC Geneva), R. Pell\`o (LAOMP Toulouse), A. Pollo (LAM Marseille), L. Pozzetti  (INAF-OA Bologna, M. Radovich (INAF-OAC Naples), M. Bondi (INAF-IRA Bologna), S. de la Torre (LAM, Marseille), Y. Mellier (IAP Paris), P. Merluzzi, S. Temporin (INAF-OA Brera), D. Vergani (INAF-IASF Milano), C.J. Walcher (LAM, Marseille)}
}
\address{$^1$LAM (UMR 6110), CNRS-Universit\'e de Provence, BP8, 13376 Marseille Cedex 12, France \\$^2$INAF-Osservatorio Astronomico di Bologna, via Ranzani 1, 40127 Bologna, Italy}
\maketitle
\abstracts{The VIMOS VLT Deep Survey (VVDS) is a unique $I$-selected
  spectroscopic sample to study galaxies all the way from $z=5$ to
  $z=0$. We recapitulate the first results about the evolution of the
  galaxy populations as a function of type, morphology, environment
  and luminosity.}

\section{Introduction}
Galaxy redshift surveys are outstanding tools for observational
cosmology because they produce large sample of {\it standard} galaxies at
different cosmic epochs. Redshift acquisition has undergome tremendous
progress thanks to advances in technology, and redshift surveys appear
nowdays as routine. Even though they may look simple at face-value,
the strategy of a survey and the galaxy-selection criteria have
crucial impacts on the interpretation of results. Since galaxies are
directly observable point-like tracers of dark matter halos, they
represent only the tip of the iceberg of what drives the evolution of
the Universe. Hence interpretation of these surveys via
model-dependent approches provide fundamental insights into galaxy
evolution and formation.

The construction of various distribution functions in different
redshift bins leads to the des\-cription of the galaxy populations
through cosmic time.  In this process a key-point is to define and
quantify the completeness of sources so we can compare galaxy samples
both at different redshift bins within a single survey, and with other
observed or simulated surveys (see discussions in Tresse, 1999). The
luminosity function, $\phi(L)$, that is the distribution of the
comoving number density of galaxies as a function of their intrinsic
luminosity, is a fundamental measurement to quantify the evolution of
the galaxy populations. Moreover, integrated quantities such as the
luminosity density, $\int\phi(L)LdL$, are little dependent on the
individual evolutionnary histories, and are solid links to the
underlying processus of galaxy evolution and formation, such as the
star formation rate (SFR) history or the stellar mass assembly.
Nevertheless, because galaxy evolution involves a complex interplay of
galaxy morphology, color, SFR, mass accretion and environment, it is
necessary to acquire large samples to trace the galaxy populations
sorted out according to these parameters. These galaxy characteristics  are supplied by the
new generation of redshift surveys, like the VVDS ({\it
  www.oamp.fr/virmos/}), over the same regions of sky as broad-band or
follow-up observations.

\section{The VIMOS VLT Deep Survey}
\begin{figure}
\vspace{-1.9cm}
\centerline{\psfig{figure=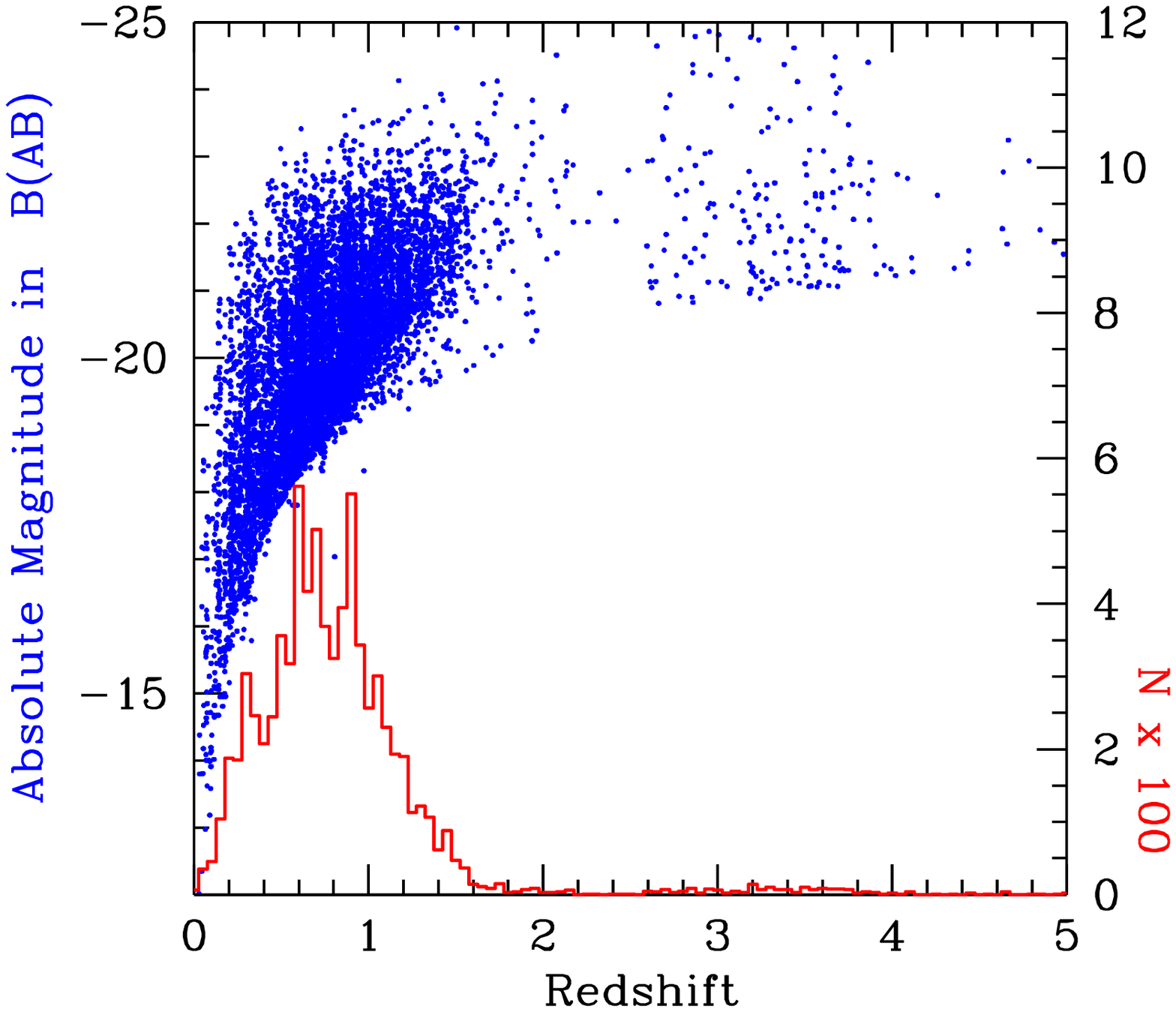,height=8cm}\hspace{-1.cm}\psfig{figure=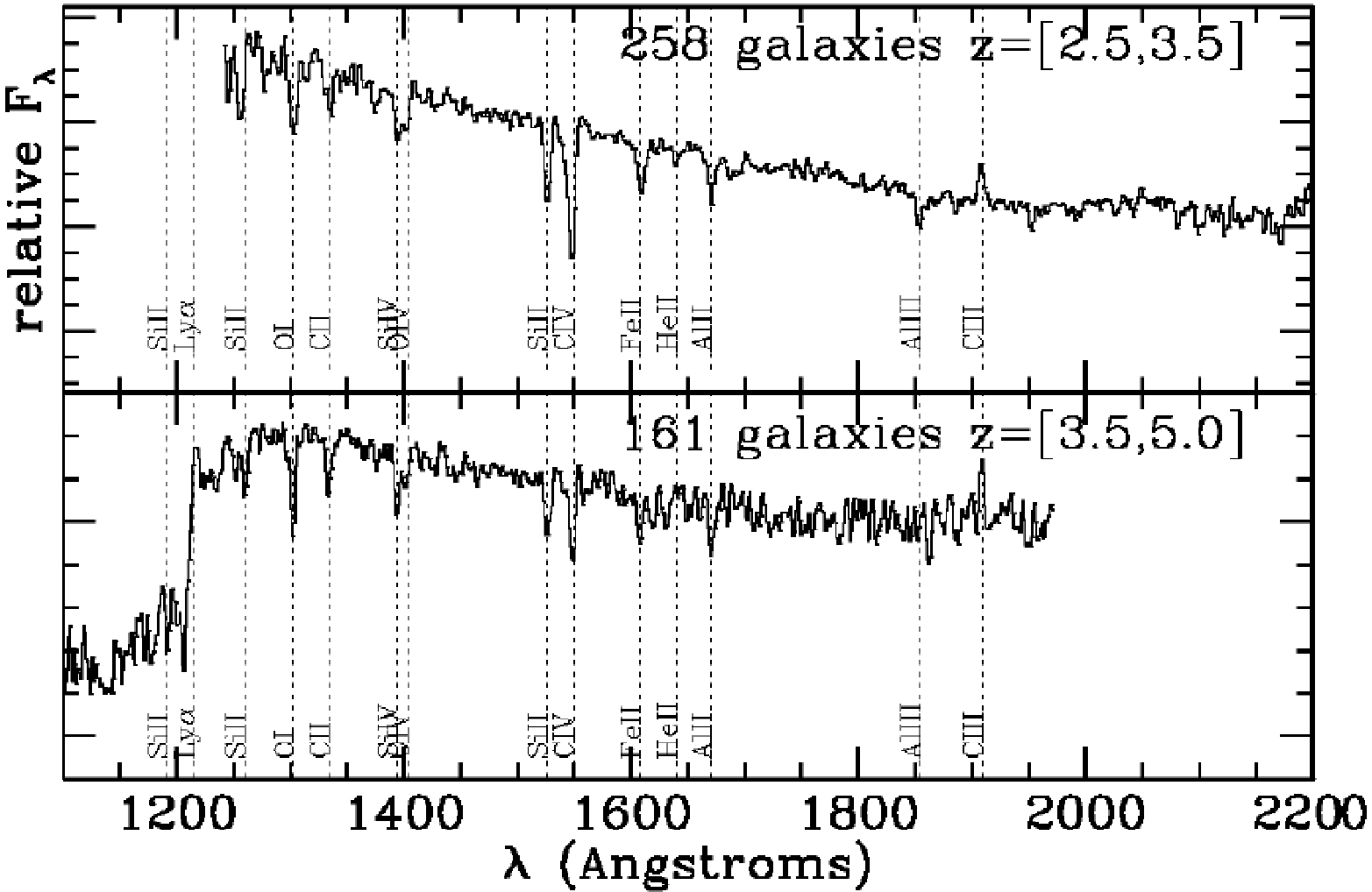,height=5.5cm}}
\caption{From left to right. (a)~The histogram of redshifts and the distribution of galaxies in the M(B$_{AB}$)$-$redshift diagram limited by $17.5\le I_{AB}\le 24$ and by the excellent efficiency to measure redshifts at $z<2$ and $z>2.7$. (b)~High-redshift VVDS spectra averaged in the redshift intervals $2.5<z<3.5$ and $3.5<z<5.0$ (from Le F\`evre et~al., 2005b). \label{fig1}}
\end{figure}
The VVDS redshift survey is based primarily on observations with the
8.2m ESO-VLT Melipal telescope in Paranal, Chili. It consists of
spectra of over $10^5$ faint sources covering $\sim10$~deg$^2$ in five
regions of sky. The survey is divided into three selection functions
with sources selected to have $17.5 \le I_{AB} \le 24.0$ (VVDS-Deep),
$17.5 \le I_{AB} \le 22.5$ (VVDS-Wide), and $I_{AB} \le 25.75$
(VVDS-Ultra Deep). Sources have been targeted on the sole criterion of
an $I_{AB}$ flux limit. No pre-selection has been applied in term of
colors, sizes, photometric redshifts, or peculiar sources.  We present
results from the first epoch observations (VVDS-Deep) obtained in two
fields of view, VVDS-02 and VVDS-CDF (see details in Le F\`evre et
al., 2004; Le F\`evre et al., 2005a). The sample is composed of 11564
spectra over 2200 arcmin$^2$ of sky area observed in UBVRI broad-band
photometry. The measured redshifts are at $0<z<5$ with an average
redshift of $0.76$ (and a median of $0.9$), and with a 1$\sigma$
accuracy of the $z$ measurement of $0.00009$ (see Fig.~\ref{fig1}a).
We have classified 79\% (7840 galaxies, 751 stars, 71 QSOs), and 14\%
(1500 galaxies, 80 stars) of spectra with confidence levels at
$>81$\%, and within [$48-58$]\%, respectively, and 7\% of spectra have not 
been identified. The most amazing discovery was the high-quality
spectra of galaxies at high redshifts, which were not expected (see
Fig.~\ref{fig1}b). We emphasize that we have obtained an excellent
efficiency for determing redshifts at $z<2$ and at $z>2.7$.  At
$2<z<2.7$ reliable spectral features are difficult to detect, and
observations extending further to the blue or into the near-infrared
are required to fill this gap with more redshifts.  Absolute magnitude
measurements are optimized accounting for the full information given
by the multi-band photometric data in a way which minimizes the
dependence to the templates used to fit the observed colors (Ilbert et
al., 2005). We adopt the set ($\Omega_M$, $\Omega_{\lambda}$, $h$) =
(0.3, 0.7, 0.7) for the cosmological parameters.

\section{The Evolution of Galaxy Population Emissivities}
\begin{figure}
\centerline{\psfig{figure=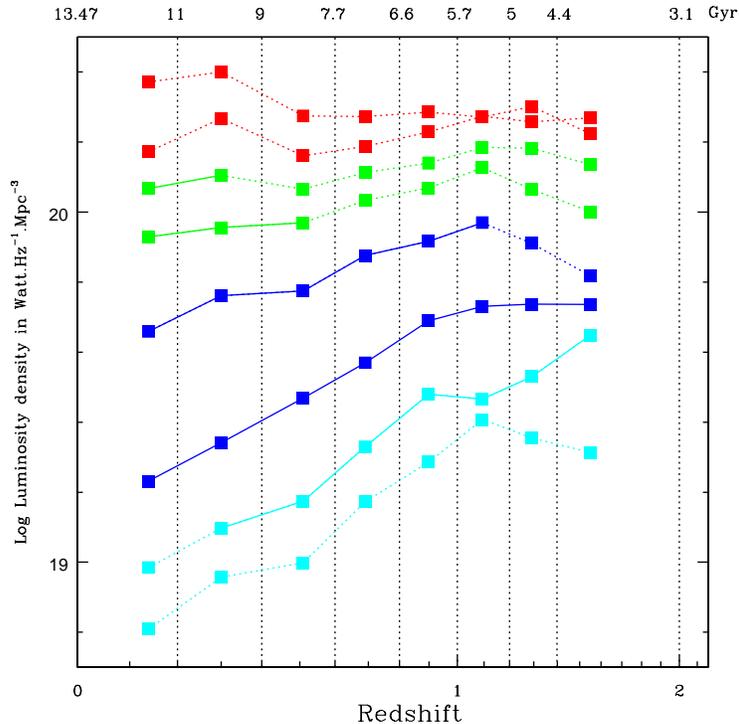,height=10cm}}
\caption{Comoving luminosity densities in the rest-frame FUV, NUV, UBVRI and K bands from bottom to top, respectively. The solid line connects points where the rest-frame band is observed in the optical. They evolve with redshift over $0.05 \le z \le 1.2$ as proportional to $(1_+z)^x$ with $x=2.05, 1.94, 1.92, 1.14, 0.73, 0.42, 0.30, -0.48$, respectively. (from Tresse et al., 2006) \label{fig2}}
\end{figure}
\subsection{Since $z=2$}
We have built the luminosity functions (LF) using the VVDS
multi-wavelength data in the rest-frame bands U-3600, B-4400, V-5500,
R-6500, I-7900 (see Ilbert et al., 2004; Ilbert et al., 2005), in the
GALEX rest-frame bands FUV-1500 (see Arnouts et al., 2005) and
NUV-2800, and in the rest-frame K-22000. The corresponding luminosity
densities (LD) are derived in summing them over all luminosities (see
Fig.~\ref{fig2}). As we do not observe the faintest galaxies (and
neither a cut-off at faint luminosities), the latter estimates are
extrapolations of the LFs using the STY estimator. The global galaxy
population exhibits a clear differential, wavelength-dependent
evolution which undergoes an upturn at redder wavelengths than the
$I$-band. This evolution is related more or less directly to the very
different stellar populations which dominate a given rest-frame band.
Although error bars are still large, most LDs display a transition at
$z\simeq1.1$ in the evolutionary tendency. Over the last 8.5~Gyrs, the
SFR-related LD(FUV) drops by a factor~4 while the stellar
mass-related LD(K) increases by a factor~1.3 in the last 4.5~Gyrs.
It might be evidence for recent merger events, but which should
produce little star formation.
\subsection{Since $z=5$}
\begin{figure}
\vspace{-3cm}
\centerline{\psfig{figure=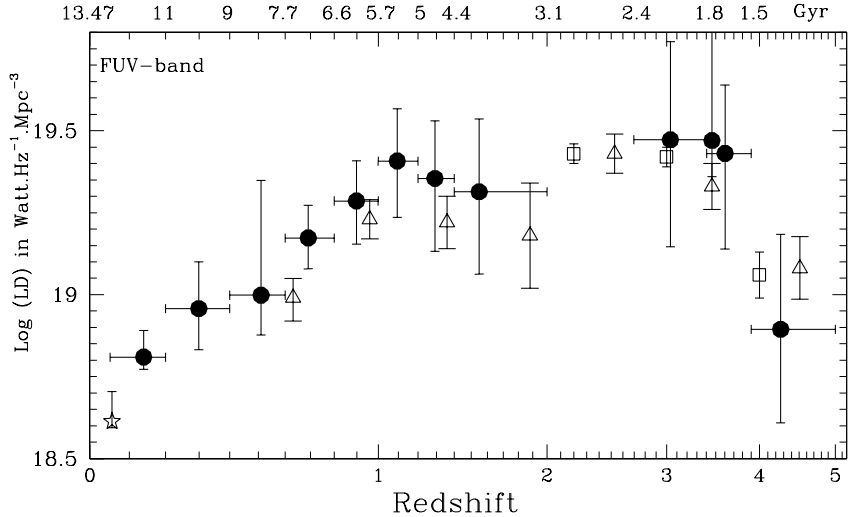,height=12cm,width=12cm}}
\caption{Comoving luminosity densities in the rest-frame FUV from $z=0$ to $z=5$. The plain circles represent the VVDS data, the open star represents the GALEX-2dFGRS local point (Wyder et al., 2005), the open triangles represent data from the FORS Deep Field (Gabasch et al., 2004) and the open squares represent data from the Keck Deep Fields (Sawicki et al., 2006). (from Tresse et al., 2006)\label{fig3}}
\end{figure}

The VVDS selection function provides a sample of high-$z$ galaxies
which have not been pre-selected given fixed criteria. It has enabled
to discover between 1.6 to 6.2 times more luminous galaxies at
$z\sim3$ (see Le F\`evre et al., 2005b; Paltani et al., 2006).  Within
a single survey, it makes possible to compare our galaxy populations
all the way from $z=5$ to $z=0$. We begin to see structures in the
global emissivity evolution, such as the several up-and-down phases
through cosmic time of the global rest-frame LD(FUV) (see
Fig.~\ref{fig3}). In particular, from $z=5$ to $z=3.4$ it increases by
at most a factor $\sim3.5$. From $z=3.4$ to $z=1.2$ it globally
decreases by a factor~1.2, with a potential decline by a factor
$\sim1.4$ from $z=3.4$ to $z=2$, and an increase by a factor~$\sim1.3$
from $z=2$ to $z=1.2$.  From $z=1.2$ to $z=0.05$ it declines steadily
by a factor~4.

\section{The Evolution per Galaxy Populations}
\begin{figure}
\centerline{\psfig{figure=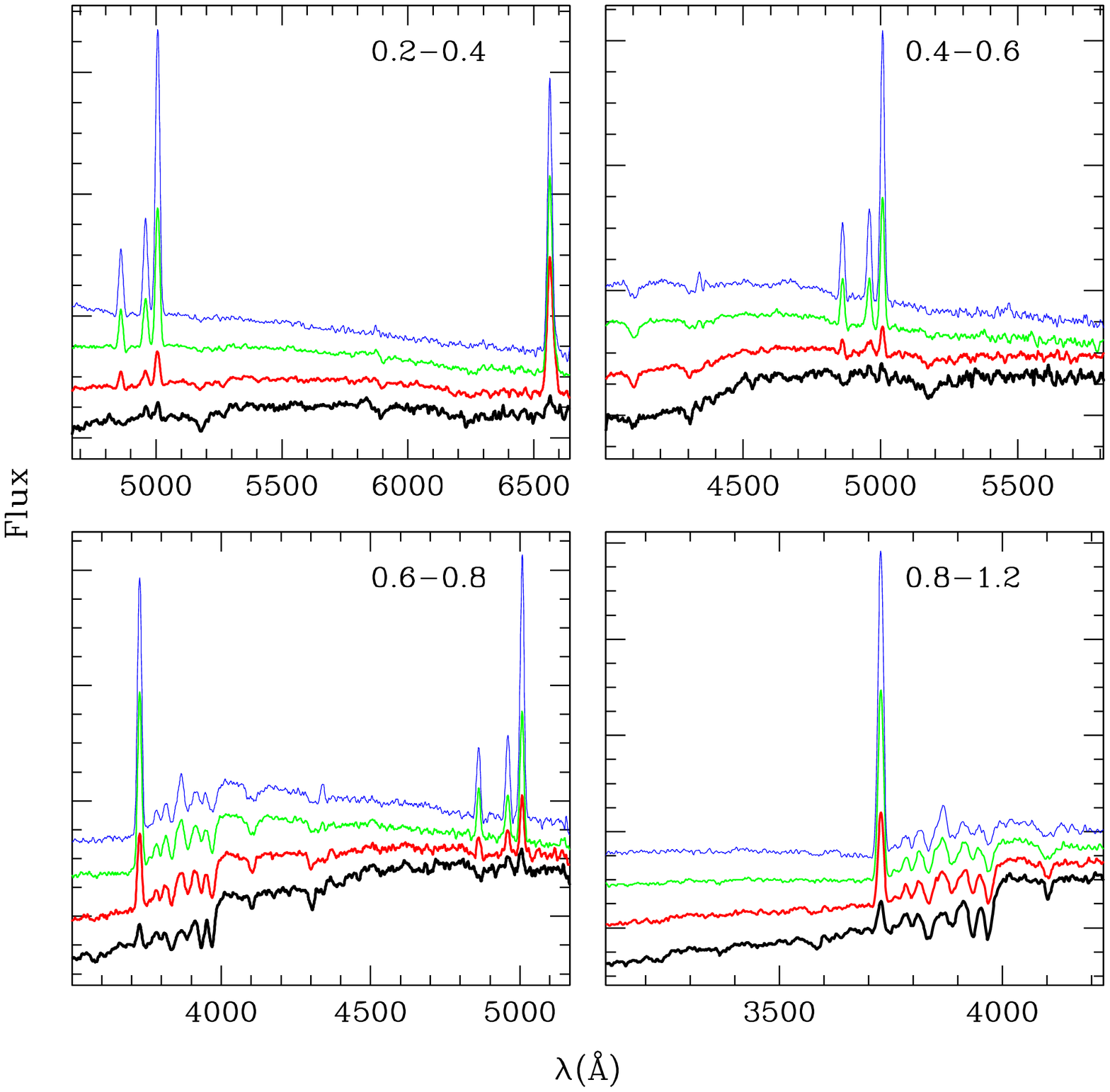,height=8cm}\psfig{figure=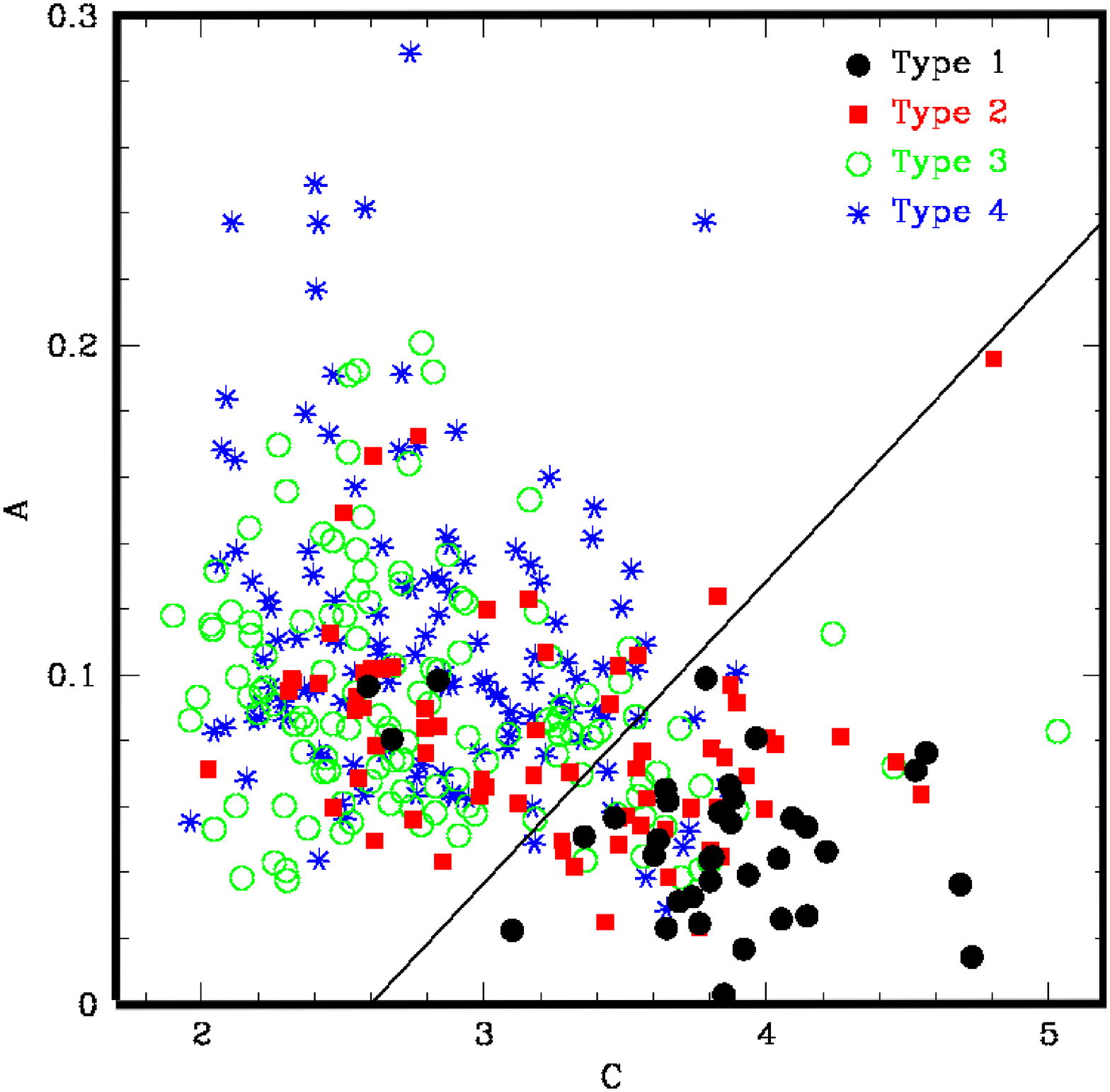,height=8cm}}
\caption{From left to right (a) Co-added VVDS spectra of the four VVDS galaxy types displayed in various redshift bins. Within a panel, the top/bottom spectra with the strongest/smallest emission lines display the latest (Type 4)/earliest (Type 1)  VVDS type. (from Zucca et al., 2006) (b) Concentration and assymetry parameters of the VVDS-CDF galaxies observed with HST-ACS images. The line separates the galaxies dominated by  a bulge (bottom-right) from the ones dominated by a disk (left). We find that $\sim91$($92$) percent of Type~1(Type~4) galaxies lie in the region of bulge(disk) dominated objects, respectively. (from Ilbert's PhD's thesis, 2004)  \label{fig4}}
\end{figure}
Galaxies present various features which can be regrouped into classes
related to fundamental observables, such as the morphology, the colors,
the luminosity, the environment, etc. Classes are usually set using
well-known local parameters.  Through their individual history,
galaxies might change of class through cosmic time, and thus {\it the
  comparative study of the distribution functions built at different
  redshifts enables us to trace the evolution of galaxy populations,
  but not the evolution of individual galaxies}.

\subsection{Per spectral type} 
Using the whole photometric information, we have classified our
galaxies in four VVDS spectral classes, using both their colors and
redshift, from early to late type galaxies. Our classification is thus
not model-dependent, in the sense that it does not take any color
evolution with redshift.  Each class is large enough to encompass
spectral energy distributions as both the four observed CWW spectra at
$z\sim0$, and luminosity-evolved templates (see discussion in Zucca et
al., 2006). That is the VVDS classes are globally related to the
standard spectroscopic and morphological shape of local normal
galaxies as it can be seen in Fig.~\ref{fig4}.

\begin{figure}
\centerline{\psfig{figure=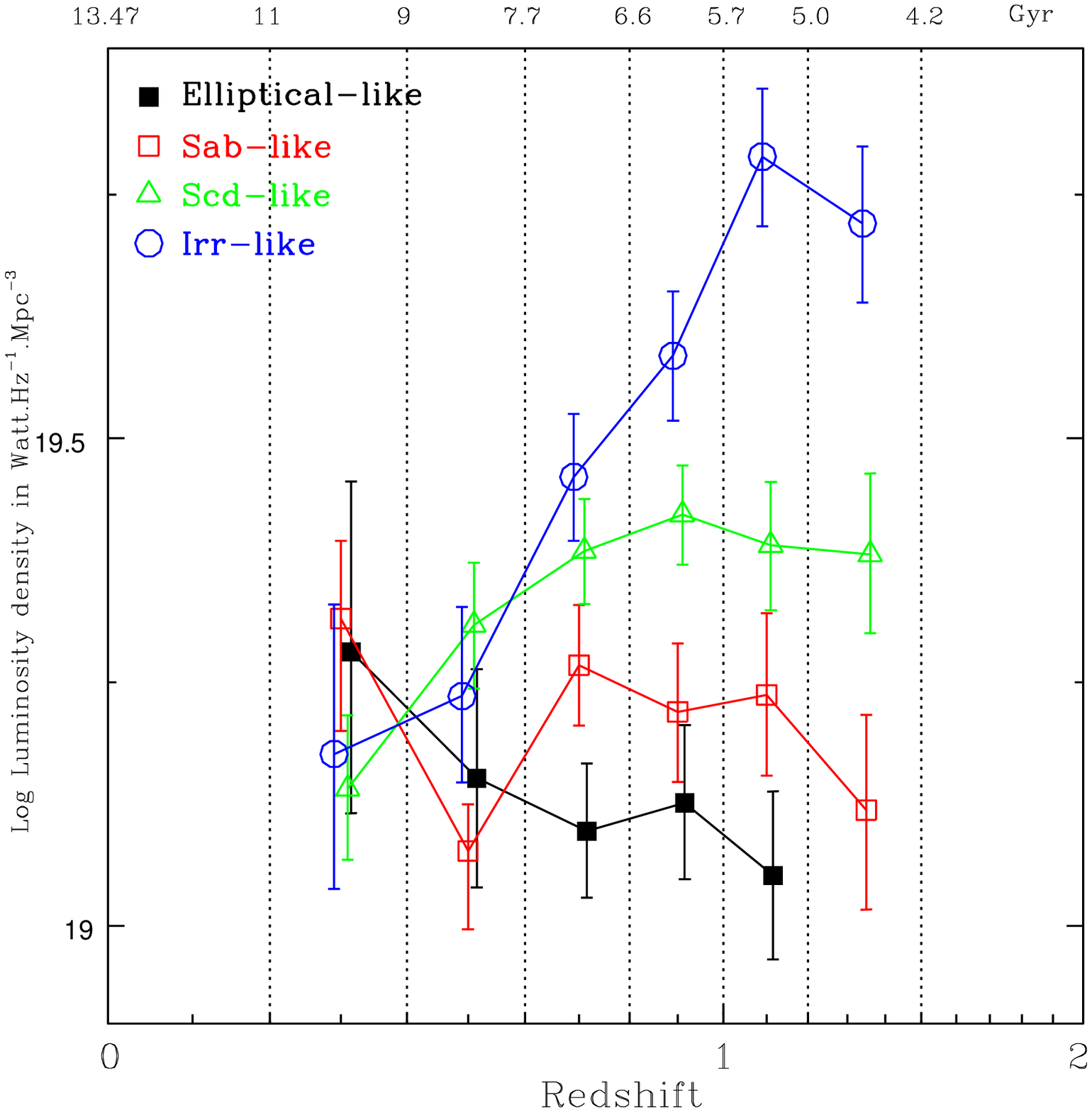,height=8cm}\psfig{figure=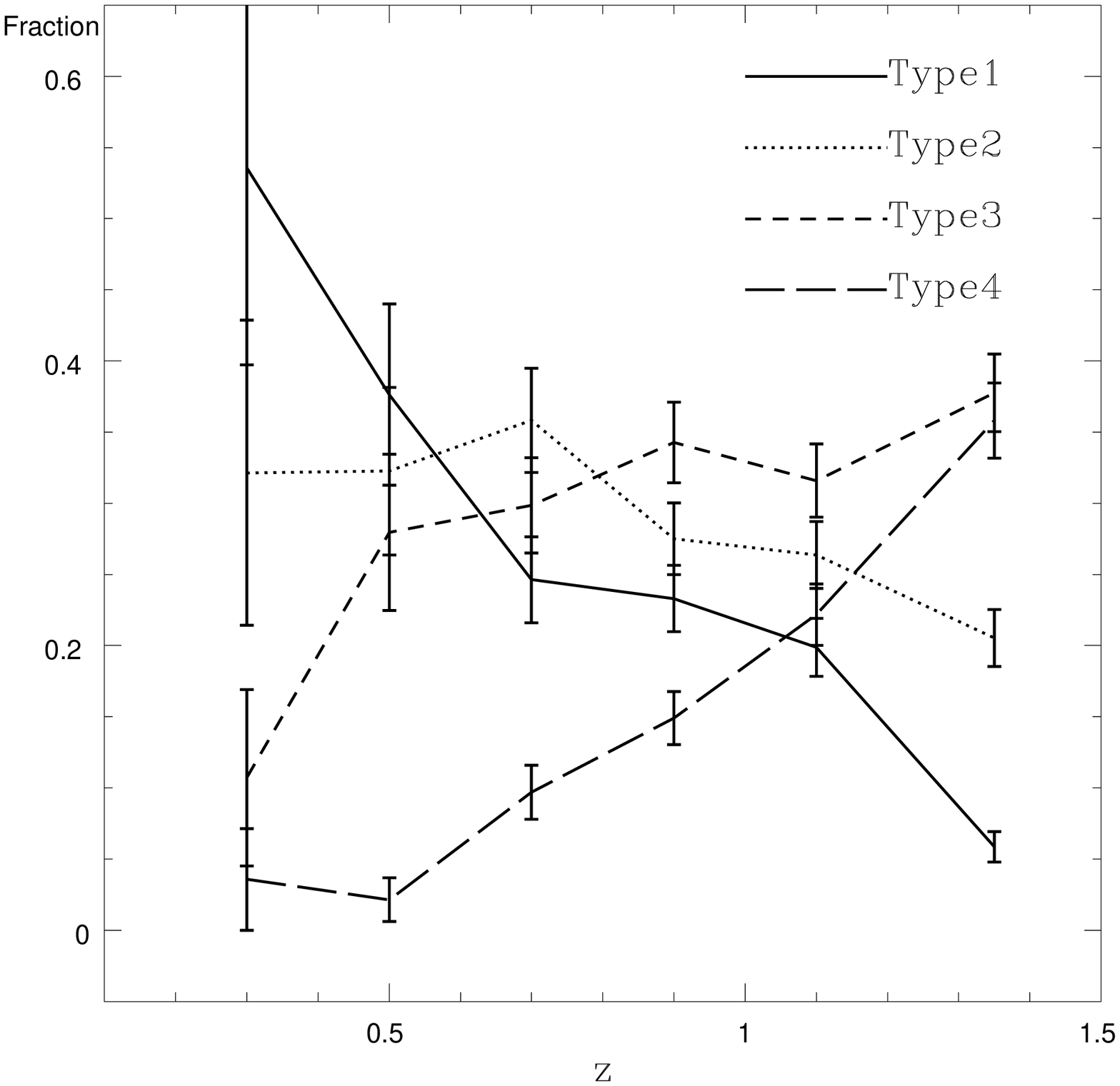,height=8cm}}
\caption{From left to right.  (a) Comoving LDs in the rest-frame B-band from early (Type~1) to late (Type~4) types. (from Tresse et al., 2006) (b) Observed fraction of bright galaxies ($M_{B_{\rm AB}} < -21.8$) from early to late types. (from Zucca et al., 2006) \label{fig5}}
\end{figure}

From Zucca et al. (2006) and Tresse et al. (2006), we find the
following results for the early and late type classes up to look-back
times corresponding to 30 percent of the current age of the Universe
(see Fig.~\ref{fig5}a).  The LF of the early-type population is
consistent with only passive luminosity evolution since $z\sim1.1$,
while the fraction of bright early-type galaxies ($M_{B}(AB) < -21.8$
mag) increases from 0.05 to 55 percent from $z=1.5$ to $z=0.2$ (see
Fig.~\ref{fig5}b). The corresponding LD(B) increases continuously by a
factor $\sim1.7$. This population suggests that luminous red galaxies
must appear at low redshifts to keep increasing the LD since this
population is faintening by 0.3 mag only.  The LF of the late-type
population undergoes a strong evolution in density and luminosity.
There is a steady decrease in volume density by a factor $\sim2$
coming from both the bright and faint parts of the LF. The fraction of
bright late-type galaxies decreases from $\sim35$ to $\sim5$ percent
from $z=1.5$ to $z=0.2$.  Thus the corresponding LD(B) decreases
markely by a factor $\sim3.5$.  This population supports a downsizing
scenario where most star formation is shifting to faint galaxies at
$z<1.2$.

\subsection{Per morphological shape} 
We have split the VVDS-CDFS galaxies in bulge- and disk-dominated
classes (see Fig.~\ref{fig4}b and Ilbert et al., 2006a). We show that
the bulge-dominated class is composed of two populations; a red and
bright galaxy populations (70\%) and a blue, compact galaxy population
(30\%).

As shown in Fig.~\ref{fig6}a, we have measured a mild evolution of the
disk-dominated population with a LD(B) decreasing by a factor $\sim
1.6$, corresponding to a $< 0.5$ mag faintening of the stellar
populations from $z=1.2$ to $z=0.05$ with no significant density
evolution.  For the red bulge-dominated population, we measure a small
increase of the LD(B) by a factor $\sim0.6$, corresponding to a
faintening of 0.2 mag and an increasing density evolution by a factor
$\sim 2$ from $z=1.2$ to $z=0.4$.  In contrast, the blue
bulge-dominated population undergoes the strongest evolution with the 
LD(B) increasing by a factor 16 in the last 8.5 Gyrs, corresponding to
a faintening of 0.7 mag.  This latter population could be the
progenitors of the local low-mass spheroidal galaxies.

\subsection{Per local environment} 
The 3D galaxy density field is reconstructed using a gaussian filter
smoothing length 5$h^{-1}$~Mpc over the VVDS-F02 field. The galaxies
have been classified in two environment classes, the under- and
over-dense environments relative to the average density contrast.  The
LF shape is strongly dependent on the large-scale environment,
suggesting that the environment has been efficiently in place at
$z>1.5$ (Ilbert et al., 2006b). Since the local measurements do not exhibit this dependence,
our result suggests that the number of faint red galaxies increases
with cosmic time in over-dense environment, consistent with the
evolution of the color-environment relationship described in Cucciati
et al. (2006).

\begin{figure}
\centerline{\psfig{figure=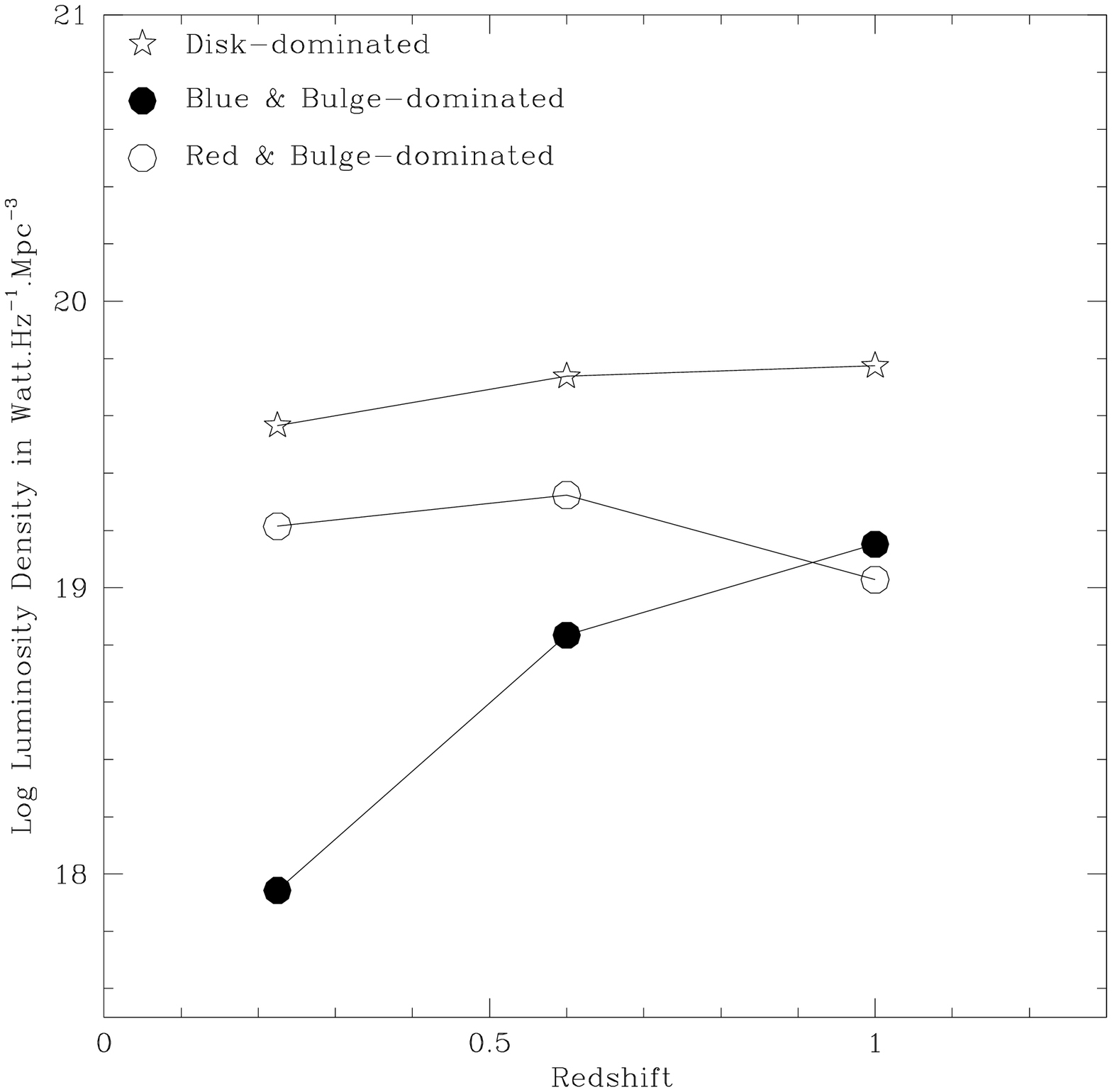,height=8cm}\psfig{figure=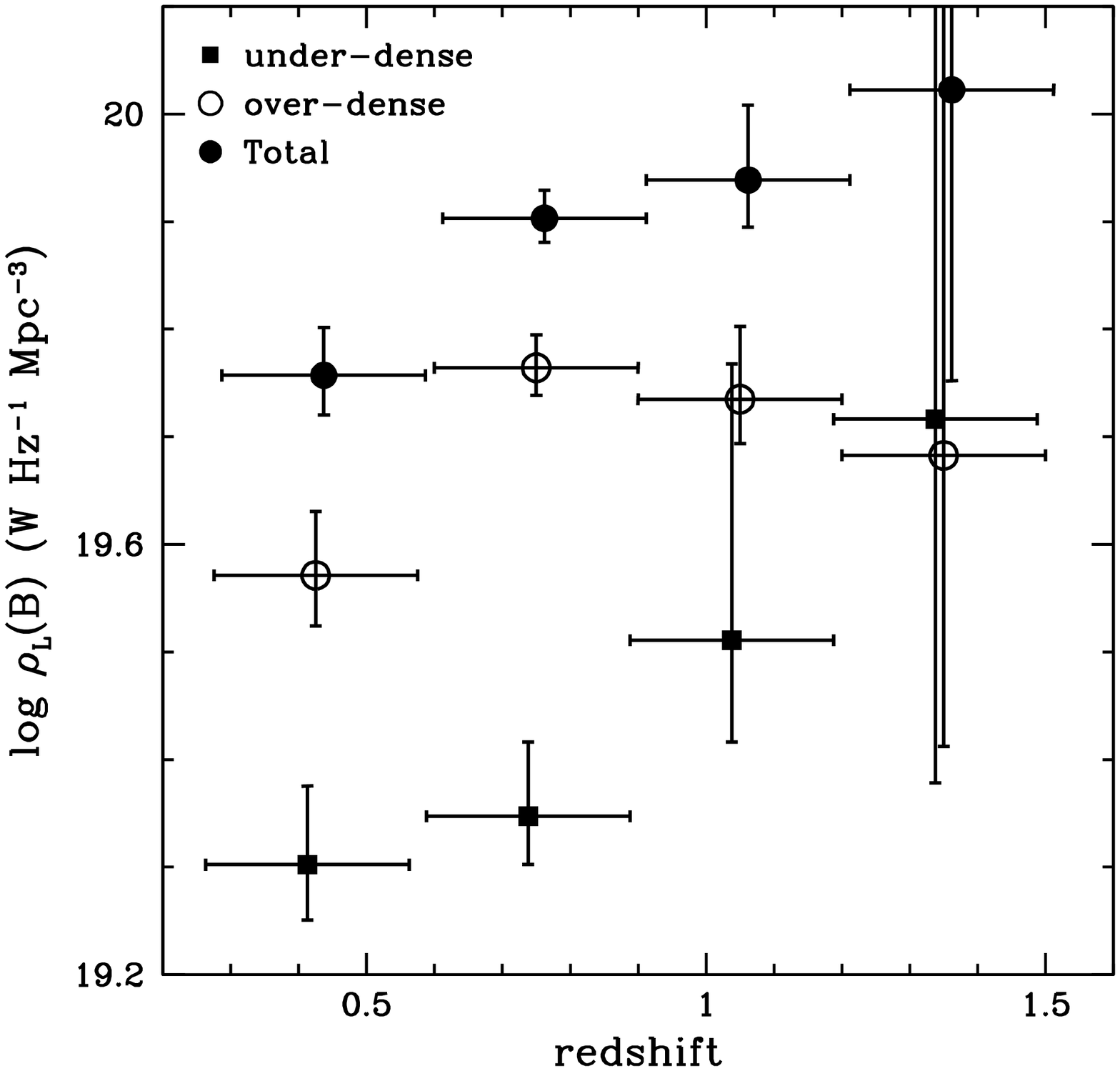,height=8cm}}
\caption{From left to right. (a) Comoving LDs in the rest-frame B-band from the disk- and bulge-dominated populations as defined in Fig.~\ref{fig4}b.  (b) Comoving LDs in the rest-frame B-band from under to over dense environments. (from Ilbert et al., 2006b)\label{fig6}}
\end{figure}

As shown in Fig.~\ref{fig6}b, from $z=1.5$ to $z=0.25$ the evolution of
the LD(B) continuously decreases by a factor $\sim2.6$ in under-dense
environments driven by faintening of 0.6 mag. It is in agreement
with the strong LD decline of the late-type polulation, and which dominates
the under-dense environments. In over-dense environments, the LD(B) increases
by a factor $\sim1.3$ from $z=1.5$ to $z=0.9$, and then decreases by a
factor $\sim1.6$. The over-dense environments are dominated by the
bright early-type galaxies and faint late-type galaxies. And thus the
over-dense environments are the place where there is a complex
interplay between the decrease of the star formation rate in late
types and the increasing fraction of bright early types.

\subsection{Per intrinsic luminosity} 
\begin{figure}
\centerline{\psfig{figure=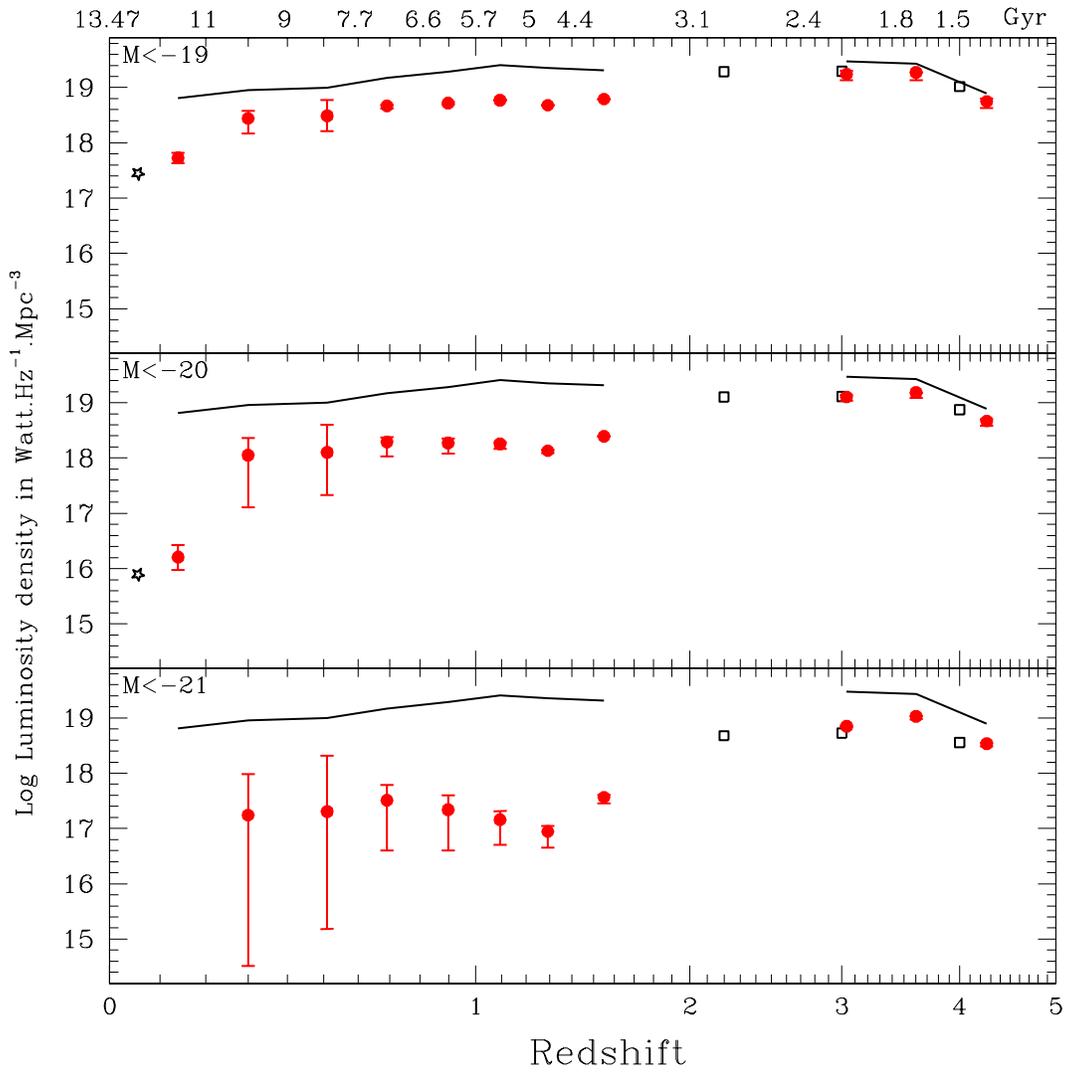,height=15cm}}
\caption{Comoving rest-frame FUV luminosity densities from $z=0$ to $z=5$ for the bright galaxy populations defined as $M_{AB}(1500A) < -19, -20$ and $-21$ mag and represented by solid circles. In each panel, the solid line connects the VVDS points of the global LD(FUV) as displayed in Fig.~\ref{fig3}. (from Tresse et al., 2006)  \label{fig7}}
\end{figure}

The SFR-related LD(FUV) is strongly luminosity-dependent as shown in
Fig.~\ref{fig7}.  The old, most luminous ($M_{AB}(1500A) < -21$)
galaxy population has exhausted its cold gas reservoir during its
early intense star formation which has occured in the early Universe
at $z\gg 4$, and since $z\simeq3.5$, i.e. 12~Gyrs, its undergoes
passive evolution as star formation cease. It creates excellent dry
candidates.  Creation of new galaxies occurs as the threshold
amplitude for forming bright galaxies decrease as described in
Marinoni et al. (2005). That is the typical $L^{*}$ of the population
created at a given redshift will decrease with decreasing redshift.
This imply that the younger, less luminous $L^*$ galaxy population
continues to efficently form stars with a large reservoir of cold gas
up to $z=0.2$. And at $z<0.2$, this later population appears to have
also exhausted its gas supply. The gas-exhaustion would favor the
evolution of morphologies toward early-type galaxies.

At $z<4$, both processes, i.e. dry mergers toward decreasing redshift
and morphologies evolving toward early-type galaxies, might contribute
to an increase of the bright early-type population by a factor
$\sim10$ to reach $\sim55$ percent of the total population from
$z=1.5$ to $z=0.2$, while the early-type population undergoes a
passive luminosity evolution.  However, it leaves little room for
mergers even dry ones from $z=1.2$ to $z=0.4$, since the mass related
LD(K) does not evolve in this redshift range, except from $z=0.4$ to
$z=0.05$ and from $z=2$ to $z=1.2$.

\section{Summary}
Within a single survey, as the VVDS, we have quantified the galaxy
population distributions since $z\sim5$. Our observed global evolution
does not seem to be in agreement with a continuous smooth decrease
from $z\sim3.5$ to $\simeq0$ as predicted by the simulations. It is
related to both the characteristics of the dominant population at a
given cosmic time and the evolution of the galaxy populations (per
type, per morphology, per environment, per luminosity).  The picture
is globally consistent with a downsizing scenario for the star
formation rate in $L^{*}$ galaxies, while the dwarf population
undergoes density evolution. The blue compact galaxy population is the
class which undergoes the strongest evolution than any other galaxy
class over the last 8.5~Gyrs. The red, spheroid galaxy population is
the class which undergoes the strongest positive density evolution
over the last $\sim 5$~Gyrs. From $z\simeq0.4$ to today, the
emissivity of the Universe is increasingly dominated by a dust
deficient galaxy population, in constrast with the more distant 
Universe dominated by dust rich, star forming galaxies.

\end{document}